\documentclass[12pt]{iopart}

\newcommand{\ket}[1]{\left| #1\right\rangle}

\newcommand{\eeqref}[1]{Eq.~(\ref{#1})}

\usepackage{graphicx}           
\graphicspath{{Images/}}

\begin{document}

\title{Multimode electromagnetically-induced transparency on a single atomic line}

\author{Geoff Campbell, Anna Ordog, A. I. Lvovsky}
\address{Institute for Quantum Information Sciences, University of Calgary, Alberta T2N 1N4, Canada}

\date{\today}


\date{\today}

\begin{abstract}
We experimentally investigate electromagnetically-induced transparency (EIT) created on an inhomogeneously broadened $5S_{1/2}-5P_{1/2}$ transition in rubidium vapor using a control field of a complex temporal shape. A comb-shaped transparency spectrum enhances the delay-bandwidth product and the light storage capacity for a matched probe pulse by a factor of about 50 compared to a single EIT line [D. D. Yavuz, Phys. Rev. A \textbf{75}, 031801 (2007)]. If the temporal mode of the control field is slowly changed while the probe is propagating through the EIT medium, the probe will adiabatically follow, providing a means to perform frequency conversion and optical routing.

\end{abstract}

\maketitle

\section{Introduction}
The control of light using electromagnetically induced transparency (EIT) \cite{EIT_review} has become a key tool in optical quantum and classical information processing. Potential applications include all optical buffers \cite{Burmeister200810} and routers \cite{RATOS1,RATOS2}, quantum gates \cite{GNL} and quantum information storage \cite{Storage}. EIT-based methods of delaying and storing large bandwidth pulses are however limited by the contrast of the transparency line. Specifically, the product between the bandwidth of the probe (signal) pulse and the maximum achievable delay cannot exceed $\approx \sqrt{d}$ where $d$ is the optical depth \emph{outside} the EIT window \cite{Novikova_07}. A similar dependence, namely $ \approx \sqrt{d}/3$ \cite{Nunn,AFC} applies to EIT-based optical memory. Although the optical density can be increased through a higher atomic number density or a longer cell, this leads to higher losses \emph{inside} the EIT line, resulting in lower efficiency. One approach to solving this problem, by storing different Fourier components of the pulse in different geometric areas of the EIT medium, has been proven to work in principle, albeit challenging to implement practically \cite{Deng}.

Another approach, investigated in the present paper, relies on an EIT window of a complex spectral shape created in a single homogeneously or inhomogeneously broadened optical transition [Fig.~\ref{fig:spectra}(a)]. The spectra of the probe and control fields are matched so that the two-photon resonance condition for EIT is met for each component \cite{Harris93}. In this way, the delay of each component (and thus the whole pulse) is determined by the inverse width of each individual EIT window, whereas the bandwidth of the pulse is given by the spectral width of the control field mode, which can be made very large. For the experimental demonstration, we employ a ``comb''-shaped EIT window created by a train of control field pulses, described theoretically by Yavuz \cite{Yavuz}.

A further application of multimode EIT is to optical routing. Changing the spectrum of the control field during the propagation of the probe pulse causes the latter to adiabatically follow. In this way, one can arbitrarily reshape the signal pulse or change its frequency within the linewidth of the optical transition.

\section{Theory} 
\begin{figure}
 \centering{
    \includegraphics[width = 0.7\columnwidth]{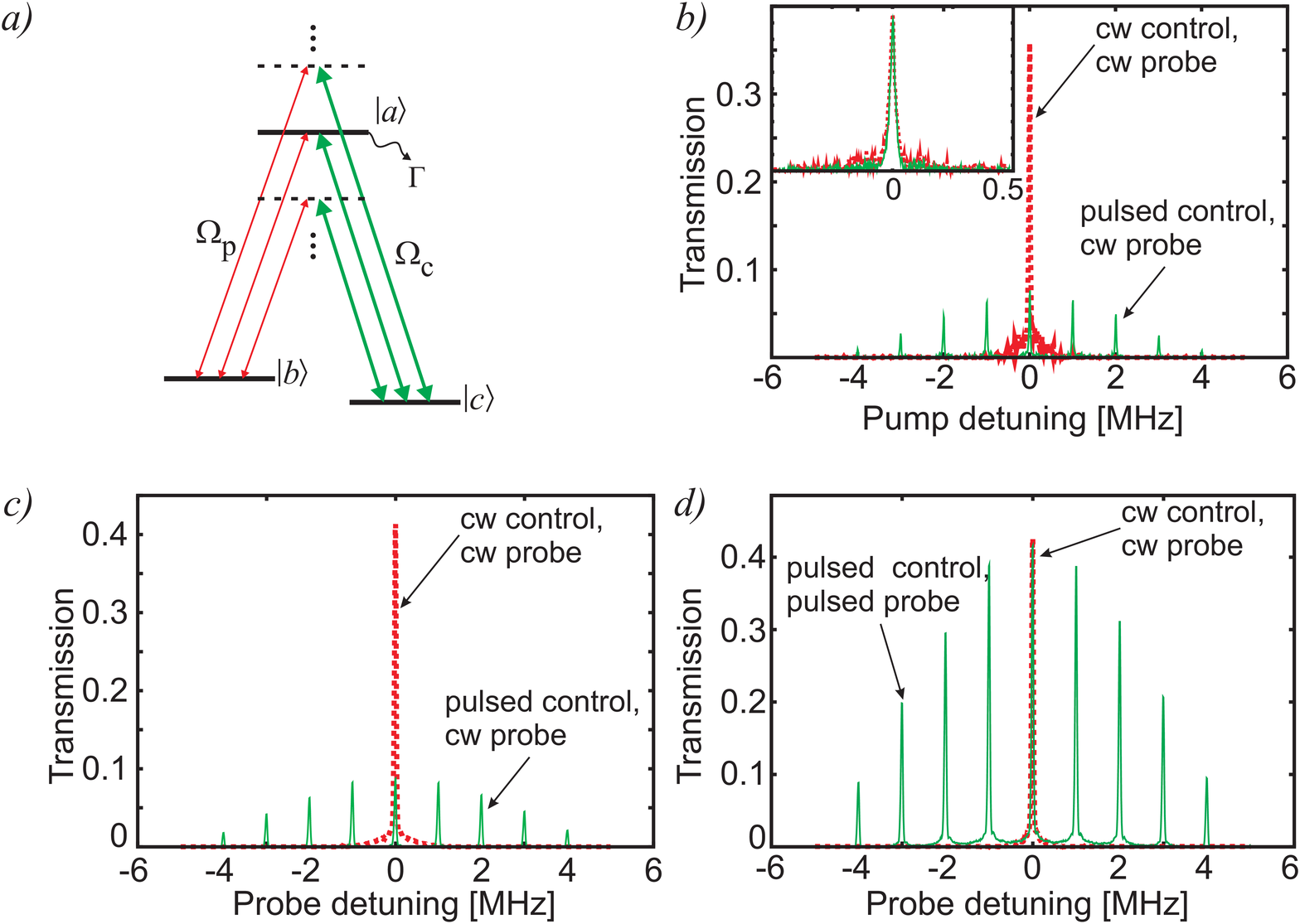}
    \caption{A comb of EIT windows. a) The atomic system employed by the protocol. b) \emph{Integrated} transmission of a cw probe field with the multi-frequency control laser scanning across the EIT resonance. When the probe is resonant with one of the comb ``teeth'', partial (20\%) EIT is present. The inset displays a $\times 10$ blowup of the horizontal axis, with the signal associated with the pulsed pump case magnified by $\times 5$, showing the EIT windows to be of the same shape. c) \emph{Spectrum} of the transmitted cw probe field with cw and pulsed control fields. Propagation through the EIT medium makes the temporal mode of the probe match that of the control. d) If the probe field is initially pulsed, i.e. matches the temporal mode of the control, there is no loss in transmission compared to the cw case. }
    \label{fig:spectra}
  }
\end{figure}

We begin by reproducing, with some generalization, the existing theory \cite{Harris93,Yavuz,Eberly94}. We study an atomic system displayed in Fig.~\ref{fig:spectra}(a). We assume the probe field to be very weak, so the stochastic wave function method can be used (i.e. the state of the atomic system can be assumed pure) and the amplitude of energy level $\ket b$ $c_b\equiv 1$. We employ the local time coordinate system \cite{Eberly94}: $\tau=t-z/c$, $\zeta=z$,
in which
the equations of motion for the probability amplitudes of other atomic levels take the form
\numparts
\begin{eqnarray}
\dot c_c &=& i \Omega^*_c c_a \label{motion1}\\
\dot c_a &=& -\Gamma/2\ c_a+i\Omega_p c_b+i\Omega_c c_c,\label{motion2}
\end{eqnarray}
\endnumparts
where dots denote partial derivatives by $\tau$, $\Omega$'s denote the Rabi frequencies of time-dependent probe and control fields, and $\Gamma$ is the spontaneous decay rate from level $\ket a$. In the adiabatic approximation (its validity limits will be discussed later), the terms containing $c_a$ in \eeqref{motion2} can be neglected and the equations are readily solved as
\numparts
\begin{eqnarray}
c_c &=& -\Omega_p/\Omega_c \label{motionsoln1}
\\
c_a &=& -i \dot c_c /\Omega^*_c.\label{motionsoln2}
\end{eqnarray}
\endnumparts

Next, we solve the propagation equation for the probe field. Because in the local time coordinates $ (1/c) \partial/{\partial t }+ \partial/{\partial z }=\partial/{\partial \zeta }$, we obtain
\begin{equation}\label{Maxwell}
c\frac{\partial \Omega_p}{\partial \zeta}=i\frac{\mu^2\omega N}{2\hbar \epsilon_0} c_a,
\end{equation}
where $\mu$ is the dipole moment of the probe transition, $\omega$ is the optical frequency of the probe and $N$ is the number density. If we rewrite this equation in terms of the amplitude $c_c$ using Eqs.~(\ref{motionsoln1}), and note that $\Omega_c$ does not depend on $\zeta$, \eeqref{Maxwell} will take convenient form
\begin{equation}\label{Maxwell1}
\frac{\partial c_c}{\partial \zeta}=- \frac{1}{\tilde V_g}\frac{\partial c_c}{\partial \tau}\ \
\textrm{with } \tilde V_g=\frac{2\hbar \epsilon_0 c |\Omega_c^2|}{\mu^2\omega N}.
\end{equation}The solution of this equation
corresponds to propagation of the atomic excitation with instantaneous group velocity $\tilde V_g$ in the local time coordinates, or with group velocity $V_g=(1/\tilde V_g+1/c)^{-1}$ in the canonical coordinates $(t,z)$.

We can now determine the adiabaticity conditions. To this end, we substitute the zero-order solution (\ref{motionsoln2}) into \eeqref{motion2} and obtain
\begin{equation}
-i\frac{\ddot c_c}{\Omega_c^*}+i\frac{\dot c_c\dot\Omega_c^*}{(\Omega_c^*)^2}=i\frac\Gamma 2 \frac{\dot c_c} {\Omega^*_c}+i\Omega_p+i\Omega_c c_c.
\end{equation}
In order for the last two terms in the right-hand side to dominate, the following conditions must be fulfilled:
\numparts
\begin{eqnarray}
|\Omega_c| T \gg 1;\label{adiabaticity1}\\
|\Omega_c|^2 TT_1 \gg 1;\label{adiabaticity2}\\
|\Omega_c|^2 T/\Gamma \gg 1\label{adiabaticity3},
\end{eqnarray}
\endnumparts
where $T$ and $T_1$ are characteristic variation times of $c_c$ and $\Omega_c$, respectively. Conditions (\ref{adiabaticity1}) and (\ref{adiabaticity3}) are inherited from classic EIT theory \cite{Storage} whereas Eq.~(\ref{adiabaticity2}) applies to the case studied here: when the optical fields and atomic coherence vary asynchronously.

Inequality (\ref{adiabaticity3}) implies that the bandwidth of the atomic ground state coherence generated by the probe pulse must not exceed the spectral width of the EIT window, $|\Omega_c|^2 /\Gamma$. This does not mean, however, that the bandwidth of the probe pulse must be equally narrow. Consider a broadband probe field with Rabi frequency of the form $\Omega_p(\zeta,\tau) = \Omega_{p0}(\zeta,\tau)f(\tau)$, $\Omega_{p0}(\zeta,\tau)$ being a slowly varying envelope and $f(\tau)$ a fast modulation. If it is injected into the EIT cell accompanied by a synchronously modulated control field
$
 \Omega_{c}(\tau) = \Omega_{c0}(\tau) f(\tau)
$
with a slowly varying $\Omega_{c0}$, the atomic ground state coherence $c_c=-\Omega_p/\Omega_c=-\Omega_{p0}/\Omega_{c0}$ will be narrowband, in spite of a wide bandwidth of the modulation function. In this case, conditions (\ref{adiabaticity1}) and (\ref{adiabaticity3}) are fulfilled as long as the envelope functions $\Omega_{p0}(t)$ and $\Omega_{c0}(t)$ are sufficiently slowly varied. The bandwidth $1/T_1$ of the modulation function, as follows from Eq. (\ref{adiabaticity2}), is then limited only by the atomic linewidth $\Gamma$.

The time averaged group velocity of the probe pulse is determined by \eeqref{Maxwell1} and depends only on the time averaged control field power. The propagation of the probe pulse envelope through the modulated EIT medium will be identical to that of an \emph{unmodulated} probe pulse $\Omega'_p(\tau)=\Omega_{p0}(\tau)$ through a single EIT window generated by unmodulated control field with $\Omega'_c(\tau)=\Omega_{c0}(\tau)$. On the other hand, the fast modulation $f(\tau)$ of the probe propagates through the medium with the speed of light in vacuum, i.e. synchronously with the control field modulation \cite{Harris93,Eberly94}.

The requirement that the fast modulation $f(t)$ of the probe field match that of the control is essential in order for the probe field to experience EIT \cite{Harris93}. Otherwise the atomic coherence $c_c=-\Omega_p/\Omega_c$ will exhibit fast variation, condition (\ref{adiabaticity2}) will not hold and the probe field will be partially absorbed in a ``Procrustean'' fashion: only the component whose temporal mode matches that of the control field will be transmitted.

This holds, perhaps counterintuitively, even in the case of inhomogeneous broadening when the sufficiently distant spectral components of the control and probe fields are in resonance with different groups of atoms. This is because the atomic dark state, $\ket b+c_c\ket c$, is shared among all atoms in the ensemble, and is insensitive of the one-photon detuning between the optical fields and the atomic transitions.

If the control Rabi frequency is slowly varied during the propagation of a matched probe pulse, the ground state coherence $c_c$ will adiabatically follow. This permits storage of broadband probe field akin to Ref.~\cite{Storage}. Furthermore, a slowly changing optical frequency of the control field will induce adiabatic following of the probe frequency. In this way, one can arbitrarily convert the optical frequency of the probe within the homogeneously or inhomogeneously broadened atomic line \cite{MoiseevHam}, yielding a useful tool for all-optical routing of quantum and classical information. Such conversion has previously been demonstrated in multilevel atoms \cite{RATOS2} and in rare-earth doped crystals \cite{RATOS3,Ham08}. An advantage of the present procedure over multilevel schemes is that the frequencies used are not restricted by the availability of appropriate transitions, but can be chosen arbitrarily within the linewidth of the excited state. The continuum of available frequencies allows the manipulation of arbitrary temporal modes, which can be used for controlling time-bin quantum information.

\section{Experiment}
\subsection{Setup}
The atomic species used in our experiment was $^{87}$Rb gas in a 12 cm vapor cell with 10 torr Neon buffer gas maintained at 50$^\circ$C in a magnetically shielded oven. The $\Lambda$ system for EIT consisted of the 5$^{2}$S$_{1/2}$, $F=1,2$ hyperfine levels as ground states and the 5$^{2}$P$_{1/2}$, $F'=1$ as the excited state. 

Two diode lasers, phase-locked at 6.834 GHz \cite{phaselock} provided the probe and control fields required for EIT. Each beam passed through an acousto-optic modulator (AOM) before propagating through the Rb cell and could be modulated or switched independently. The average control field power in most measurements was 0.6 mW. Both fields were in the same circular polarization. 

The transmitted probe was measured using heterodyne detection. The local oscillator, tuned 200 MHz from the probe, was overlapped with the fields emerging from the rubidium cell on a fast photodiode. The beat note was detected with a spectrum analyzer.

\subsection{EIT comb}
In order to generate a frequency comb of EIT windows, we switched the control field on and off with a 1 MHz frequency and a 20\% duty cycle. We then measured the transmission of the probe as a function of the two-photon detuning of the modulated control field. The spectrum analyzer was set to operate in the zero span mode with the resolution bandwidth set to 5 MHz, so the spectral structure of the transmitted probe was not resolved. We observed transmission when one of the spectral components of the modulated control field was in two-photon resonance with the probe [Fig.~\ref{fig:spectra}(b)]. However, even in that case the transparency was only partial because of the ``Procrustean'' effect described above: only the component of the temporal mode of the probe that matches that of the control field is transmitted. Because the overlap between the temporal modes of the cw probe and the pulsed control was $1/5$, the overall transmission of the probe was about $1/5$ of the transmission in the case of unmodulated fields.

The transmitted component of the probe has a comb-like spectrum, as evidenced by Fig.~\ref{fig:spectra}(c). Here the spectrum of the transmitted probe was measured by turning off the scanning of the control field and changing the resolution bandwidth of the spectrum analyzer to 30 kHz. For comparison, we measured the transmission in the case when the initial probe field was modulated similarly to the control, so their temporal modes matched each other. Under these conditions, the transmission was not compromised [Fig.~\ref{fig:spectra}(d)].

\begin{figure}
 \centering
  \includegraphics[width = 0.7\columnwidth]{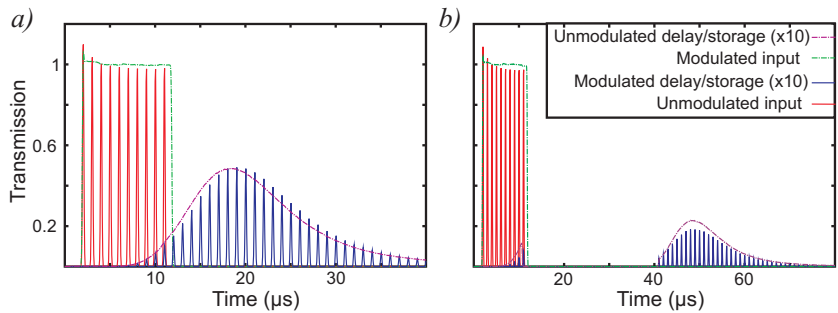}
 \caption{Delayed (a) and stored (b) unmodulated and modulated pulses with a 10$\mu$s long envelope. The envelopes of the modulated and unmodulated pulses match for equal control field powers.}
 \label{fig:delay}
\end{figure}

\subsection{Broadband delay and storage}

Broadband delay was demonstrated by modulating the probe similarly to the control field within an envelope of a duration that corresponded to the inverse EIT linewidth. Figure \ref{fig:delay}(a) shows the transmitted modulated pulse with a matched control field. For comparison, we measured the transmission of an unmodulated probe pulse within the same envelope with a cw control field of the same average power. Storage of a broadband probe pulse was similarly attained, by switching off the modulated control field and then switching it back on a short time later. For comparison, an unmodulated probe pulse was likewise stored and retrieved by means of an unmodulated control field [Fig.~\ref{fig:delay}(b)].

The envelopes of the delayed and retrieved pulses are almost identical in the modulated and unmodulated cases, indicating that ordinary and broadband pulses can be delayed and stored for the same duration. The delay-bandwidth product can hence be estimated to increase, compared to the unmodulated case, by a factor of $W/\gamma_{EIT}$ where $W$ is the bandwidth of the modulation and $\gamma_{EIT}$ is the EIT linewidth. In our experiment the modulation bandwidth is limited to $\approx$5 MHz, resulting in an approximately 50-fold increase of the delay-time bandwidth product and multi-mode storage capacity. As discussed above, the theoretical limitation of the pulse bandwidth is much higher. In order to verify this experimentally, we would however need a larger modulation bandwidth that could not be provided by the AOMs used.

\begin{figure}
 \centering
  \includegraphics[width = 0.35\columnwidth]{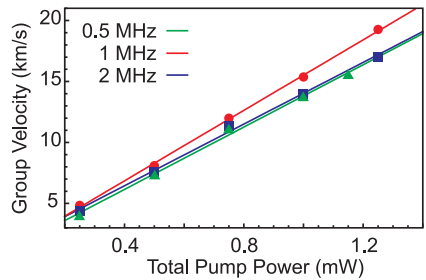}
 \caption{Group velocity as a function of control power for different modulation frequencies.  }
 \label{fig:groupvels}
\end{figure}

Figure \ref{fig:groupvels} shows the group velocity as a function of the control field power for different modulation frequencies. We find, in accordance with the theory, that the group velocity depends only on the average power of the control field across all frequency components. A similar behavior was observed when the duty cycle of the pulsed fields was varied between 10\% and 80\%.

\subsection{Frequency conversion}

\begin{figure}
 \centering{
    \includegraphics[width = 0.7\columnwidth]{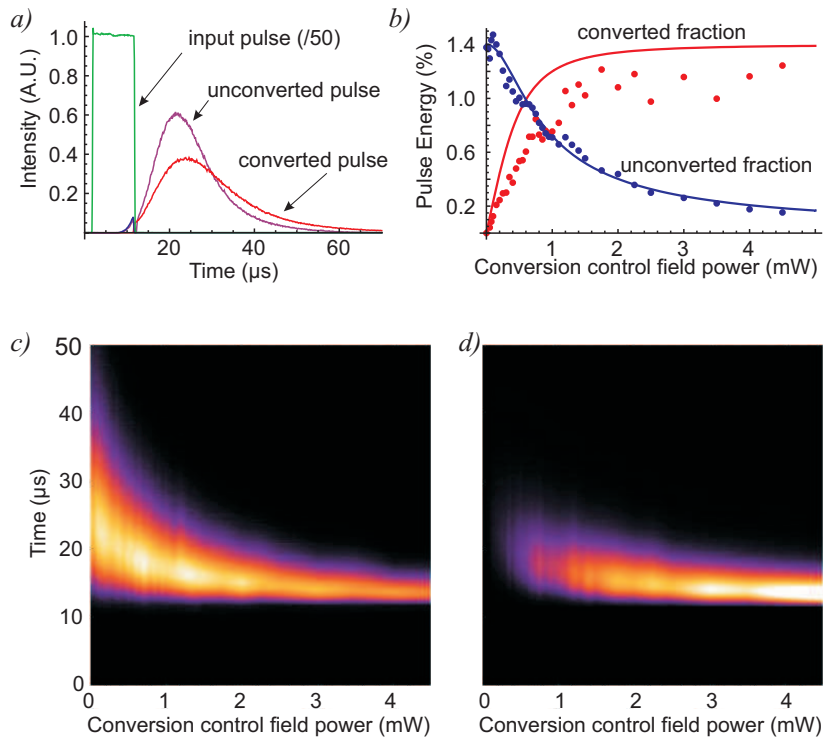}
    \caption{Frequency conversion by adiabatically changing the mode of the control field. a) A probe pulse is frequency-shifted by 160 MHz with 87\% efficiency. b) A monochromatic input probe pulse is split into frequency modes. The ratio is controlled by the ratio of control fields, which is displayed by solid lines. c) and d) The shapes of the two probe field frequency components as the second control power is varied. The group velocities of the original (c) and the converted (d) frequencies are matched.}
    \label{fig:RATOS}
  }
\end{figure}

We demonstrated adiabatic frequency conversion of the probe signal by changing the control field frequency during the propagation of a (narrowband) probe pulse through the EIT medium. To this end, we applied an additional control field, whose frequency differed from the initial control field frequency by 160 MHz, after the signal pulse entered the cell. Both control fields were well within the 550-MHz inhomogeneous line of the rubidium transition. If the initial control field was switched off during the probe propagation, the frequency of the probe emerging from the cell shifted by 160 MHz, as shown in Fig.~\ref{fig:RATOS}(a). The conversion efficiency as compared to simple delay was found to be 87\%.

If we did not turn off the initial control field after turning on the second one, a two-color pulse was generated. The ratio of powers in each frequency was equal to the ratio of respective control powers, as shown in Fig.~\ref{fig:RATOS}(b) \cite{RATOS1}. As is expected from \eeqref{Maxwell1} the group velocity of the pulse is governed only by the total power of the two control fields. The probe pulses of the two frequencies are locked together despite very different respective control field powers [Fig.~\ref{fig:RATOS}(c),(d)].

\section{Applications in quantum information technology}
Although the method reported here succeeds in increasing the delay-time-bandwidth product, its obvious limitation is the requirement that the temporal modes of probe and control must match. In application to \emph{classical} information storage, this means prior knowledge of the probe is needed in order to use the proper control field components for delay and storage. As such, the delay-\textit{information}-bandwidth is not increased and broadband storage can be viewed as storage of a single, complex temporal  mode.

Equally problematic is the application of this method to \emph{quantum} information storage. The quantum information transferred from light to the atomic ensemble is carried in the ground state coherence, quantified by the amplitude $c_c$, which plays the role of the dark-state polariton \cite{Storage} in the local time coordinates. This coherence obeys the same spectral restriction and propagates with the same group velocity as in the case of classic EIT, and hence does not possess enhanced information storage capacity. We note that a recently reported optical storage method based on rephasing of a comb-shaped inhomogeneously broadened line \cite{AFC,AFC1} does not employ a dark state and thus does not suffer from this problem.

Nonetheless, the scheme has merit for other purposes such as obtaining enhanced non-linearities: two weak pulses with matched group velocities \cite{DEIT,DEIT1} could experience a lengthened interaction time for given peak intensities using this method. Additionally, the scheme can be used for linear conversion of classical and quantum information within a broadened atomic transition. Arbitrary linear manipulation of temporal or frequency modes could be performed.

\section{Conclusion}

We have successfully implemented a method for significantly increasing the delay-bandwidth product by means of a multi-channel EIT system in a warm atomic vapour. The results were in agreement with the theoretical predictions, confirming that the delay time depends only on the average control field power, and that exactly matched Fourier components of control and probe fields are required for efficient and accurate delay and retrieval of a broadband probe pulse. An advantage of this scheme over previous EIT channelization experiments is its simplicity because it does not require spatial separation of spectral components, preparation of the medium or exceptionally large broadened linewidths.

This work was supported by NSERC, iCORE, CFI, AIF, Quantum$Works$, and CIFAR. We thank D. Yavuz, M. O. Scully, S. A. Moiseev and A. MacRae for helpful discussions.


\section*{References}

\begin{thebibliography}{10}

\bibitem{EIT_review}
M.~Fleishhauer, A.~Imamogl\v{u}, and P.~Marangos 2005
\newblock {\em Rev. Mod. Phys}  {\bf 77} 633

\bibitem{Burmeister200810}
E.F. Burmeister, D.J. Blumenthal, and J.E. Bowers 2008
\newblock {\em Optical Switching and Networking} 5 10

\bibitem{RATOS1}
J. Appel, K.-P. Marzlin, and A.~I. Lvovsky 2006
\newblock {\em Phys. Rev. A} {\bf 73} 013804

\bibitem{RATOS2}
F. Vewinger, J. Appel, E. Figueroa, and A.~I. Lvovsky 2007.
\newblock {\em Opt. Lett.}  {\bf 32} 2771.

\bibitem{GNL}
K.-P. Marzlin, Z.-B. Wang, and B.~Sanders 2006
\newblock {\em Phys. Rev. Lett.}, {\bf 97} 063901

\bibitem{Storage}
M.~Fleischhauer and M.~D. Lukin 2002
\newblock {\em Phys. Rev. A}  65 022314 

\bibitem{Novikova_07}
I.~Novikova, D.~Phillips, and R.~L. Walsworth 2007
\newblock {\em Phys. Rev. Lett.} {\bf 99} 173604

\bibitem{Nunn}
J.~Nunn, K~Reim, K.~C. Lee, V.~O. Lorenz, B.~J. Sussman, I.~A. Walmsley, and
  D.~Jaksch 2008
\newblock {\em Phys. Rev. Lett.} {\bf 101} 260502

\bibitem{AFC}
M.~Afzelius, C.~Simon, H.~de~Riedmatten, and N.~Gisin 2009 
\newblock {\em arXiv:0805.4164v3 [quant-ph]}

\bibitem{Deng}
Z.~Deng, D.-K. Qing, P.~Hemmer, C.~H.~R. Ooi, M.~S. Zubairy, and M.~O. Scully 2006
\newblock {\em Phys. Rev. Lett.} {\bf 96} 023602

\bibitem{Harris93}
S.~E. Harris 1993 
\newblock {\em Phys. Rev. Lett.}, {\bf 70} 552

\bibitem{Yavuz}
D.~Yavuz 2007
\newblock {\em Phys. Rev. A} {\bf 75} 031801 

\bibitem{Eberly94}
R.~Grobe, F.~T. Hioe, and J.~H. Eberly 1994
\newblock {\em Phys. Rev. Lett.}  {\bf 73} 3183

\bibitem{MoiseevHam}
S.~A. Moiseev and B.~S. Ham 2006
\newblock {\em Phys. Rev. A }
  {\bf 73} 033812

\bibitem{RATOS3}
H.-H. Wang, A.-J. Li, D.-M. Du, Y.-F. Fan, L. Wang, Z.-H. Kang, Y.
  Jiang, J.-H. Wu, and J.-Y. Gao 2008.
\newblock {\em Appl. Phys. Lett.} {\bf 93} 221112

\bibitem{Ham08}
B.~S. Ham 2008
\newblock {\em Phys. Rev. A} {\bf 78} 011808

\bibitem{phaselock}
J. Appel, A.~MacRae, and A.~I. Lvovsky 2009
\newblock {\em Meas. Sci. Tech.} {\bf 20} 055302

\bibitem{AFC1}
H. de~Riedmatten, M. Afzelius, M.~U. Staudt, Ch. Simon, and
  N. Gisin 2008
\newblock {\em Nature}, {\bf 456} 773

\bibitem{DEIT}
A.~MacRae, G.~Campbell, and A.~I. Lvovsky 2008
\newblock {\em Opt. Lett.} {\bf 33} 2659

\bibitem{DEIT1}
Z.-B. Wang, K.-P. Marzlin, and B.~C. Sanders 2006
\newblock {\em Phys. Rev. Lett.} {\bf 97} 063901

\end{thebibliography}
\end{document}